\documentclass[aps,twocolumn,pra,tightenlines,floatfix,showpacs]{revtex4}
\usepackage[dvips]{graphicx}
\usepackage[english]{babel}
\usepackage{amsmath}
\usepackage{amssymb}
\usepackage{times}
\newcommand{\bs}{\begin{split}}
\newcommand{\es}{\end{split}}
\newcommand{\be}{\begin{equation}}
\newcommand{\ee}{\end{equation}}
\newcommand{\ba}{\begin{eqnarray}}
\newcommand{\ea}{\end{eqnarray}}
\newcommand{\ek}{\epsilon_{\mathbf{k}}}
\newcommand{\Ek}{E_{\mathbf{k}}}
\newcommand{\uk}{u_{\mathbf{k}}}
\newcommand{\vk}{v_{\mathbf{k}}}
\newcommand{\sumk}{\sum_{\mathbf{k}}}

\newcommand{\Omegaq}{\Omega_{\mathbf{q}}}

\begin{document}

\title{Finite temperature effects in trapped Fermi gases with population
  imbalance}

\author{Chih-Chun Chien, Qijin Chen, Yan He, and K. Levin}

\affiliation{James Franck Institute and Department of Physics,
 University of Chicago, Chicago, Illinois 60637}

\date{\today}

\begin{abstract}
  We study the finite temperature $T$ behavior of trapped Fermi gases as
  they undergo BCS-Bose Einstein condensation (BEC) crossover, in the
  presence of a population imbalance.  Our results, in qualitative
  agreement with recent experiments, show how the superfluid phase
  transition is directly reflected in the particle density profiles. We
  demonstrate that at $ T \neq 0$ and in the near-BEC and unitary
  regimes, the polarization is excluded from the superfluid core.
  Nevertheless a substantial polarization fraction is carried by a
  normal region of the trap having strong pair correlations, which we
  associate with noncondensed pairs or the ``pseudogap phase''.
\end{abstract}

\pacs{03.75.Hh, 03.75.Ss, 74.20.-z \hfill \textsf{\textbf{cond-mat/0605684}}}

\maketitle

Recent work \cite{ZSSK06,PLKLH06,ZSSK206} on trapped atomic Fermi gases
with population imbalance has become particularly exciting.  With the
application of a magnetic field, these systems exhibit a continuous
evolution \cite{Leggett,Chen2,ourreview}from BCS to Bose-Einstein
condensation (BEC).  Not only are these gases possible prototypes for
condensed matter systems \cite{SR06,YD05} in the presence of a magnetic
field - Zeeman coupling, but they may also be prototypes for particle
and nuclear physics systems \cite{LW03,FGLW05}.  These pioneering
experiments have been done so far by two experimental groups
\cite{ZSSK06,PLKLH06}.  And what differences are present appear to lie
more in the interpretation than in the actual data.

There are a number of key experimental observations which we now list.
(i) Both groups have observed that the trap profiles are characterized
by a central core of (at most) weakly polarized superfluid, surrounded
by a \textit{normal region where the bulk of the polarization is
  contained}. (ii) The normal region appears to consist of overlapping
clouds of both spin states (``normal mixture''), followed at the edge of
the cloud by a region consisting only of the majority component.  There
is not complete agreement \cite{ZSSK06,PLKLH06,ZSSK206} on whether the
normal-superfluid boundary is sharp which would correspond
to some form of phase separation.

These population imbalance experiments have been done
\cite{ZSSK06,ZSSK206} in conjunction with other measurements (vortex
excitations and magnetic field sweeps) which establish the presence and
the location for superfluid condensation.  (iii) Even more recently
\cite{ZSSK206} it has been demonstrated that the superfluid phase
transition at $T_c$ can be directly reflected in changes in the shape of
the clouds. (iv) Important for the present purposes is the fact that
\cite{ZSSK206} there are strong interaction effects within the normal
region of the cloud.

The goal of the present paper is to address the four points [(i)-(iv)]
listed above through a finite temperature theory of BCS-BEC crossover in
the presence of population imbalance within a trap.  A related study of
the homogeneous system was presented earlier \cite{CCHL06}.  What is
unique to our work is the capability of separating in a natural way the
condensed from noncondensed pair contributions to the trap profile. The
difficulty in making this separation lies in the fact that (except at $T
=0$) the presence of a fermionic excitation gap, is \textit{not} a
signature of phase coherent superconductivity.  We also present
calculations of $T_c$ in a trap and show how the general shape of the
profile changes below and above $T_c$, unlike what is found
experimentally and theoretically \cite{JS5} in the absence of
polarization.  In a related fashion, we examine the noncondensed pair
states in the trap and determine to what extent they differ from a free
gas mixture of the two spin states.

Our principal findings are at general $ T \neq 0$ and for the unitary
and near-BEC regimes, (a) the superfluid core seems to be robustly
maintained at nearly zero polarization.  (b) The mixed normal region,
carries a significant fraction of the polarization within a non-superfluid
state having strong pair correlations.  Indeed, experiments suggest
\cite{ZSSK206} that ``even in the normal state, strong interactions
significantly deform the density profile of the majority spin
component".  Here we interpret these correlations as noncondensed pairs
which have no counterpart at $T=0$ and which are associated with an
excitation gap (``pseudogap") in the fermionic spectrum.  Finally, (c)
in the course of making contact with points (i)-(iv) listed above we
show good qualitative agreement with experiment.

Because we restrict our attention to condensates (and their pseudogap
phase counterparts) with zero momentum ($q_0 =0$) pairing, we do not
explore those regimes of the phase diagram corresponding to the lowest
temperatures, and highest polarizations.  Recent very nice theoretical
work based on the Bogoliubov-de Gennes (BdG) approach
\cite{Kinnunen,Machida2}, has shown that in the \textit{ground} state at
unitarity the $q_0 \neq 0$, Fulde-Ferrell-Larkin-Ovchinnikov (FFLO)
state \cite{FFLO} must be incorporated.  Importantly, the polarization
in this state at $T=0$ appears at the edge but \textit{within the
  condensate} \cite{Machida2}.  Fortunately, the present work provides a
good indication of where the FFLO phase will enter, since it occurs when
the $q_0=0$ phase is found to be unstable \cite{Machida2,Tsinghuagroup}.
With the finite $T$ required by experiments, it is quite possible that
the FFLO phase presents itself, primarily in the form of noncondensed
pairs, which lie on the perimeter of the condensate.

\begin{figure*}[t]
%\centerline{  \includegraphics[width=4.3in,clip]{Iap15_nocap_2.eps}
%\centerline{  \includegraphics[height=2.56in,clip]{Iap15_nocap_2.eps}
%~  \includegraphics[height=2.6in,clip]{col_Iap15_2.eps}}
\centerline{\includegraphics[height=2.6in,clip]{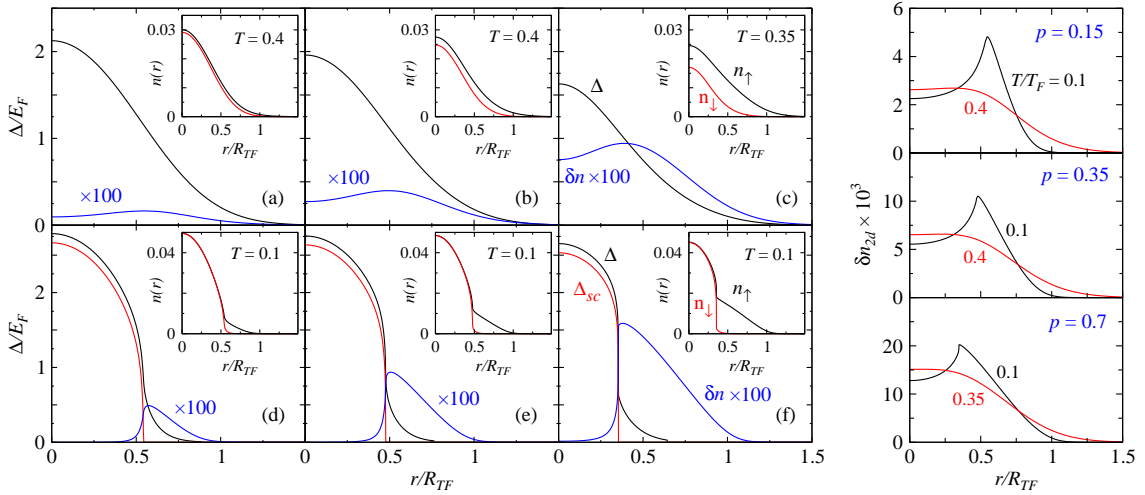}}
\caption{(Color online) Spatial distribution of the excitation gap
  $\Delta(r)$ (black) and order parameter $\Delta_{sc}(r)$ (red, main
  figures) and density $n_\sigma(r)$ (insets) at $1/k_Fa=1.5$ for the
  majority (black) and minority (red) fermions at different
  polarizations ($p=0.15$, $0.35$, and $0.7$ from left to right) above
  (upper row) and below (lower row) $T_c$. Here $T_c/T_F\approx 0.36$
  ,$0.35$, and $0.31$, respectively.  The density difference $\delta
  n(r)$ is shown in blue in the main figures, sharing the same vertical
  axis as $\Delta(r)$. The temperatures for the upper row are $T/T_F =
  0.4$, 0.4, and 0.35, respectively, and for the lower row $T/T_F =0.1$.
  Shown on the far right is the difference in column density, $\delta
  n_{2d} (r)$, for the three polarizations above (red) and below (black)
  $T_c$.  Here $E_F = \hbar^2 k_F^2/2m = k_B T_F$ is given by the Fermi
  energy for an unpolarized, noninteracting Fermi gas with the same
  total number $N$ at $T=0$, and $R_{TF} \equiv \sqrt{2E_F/m\omega^2}$
  is the Thomas-Fermi radius. The units for $n$ and $\delta n_{2d}$ are
  $k_F^3$ and $k_F^2$, respectively.}
\label{fig:1}
\end{figure*}

The value of the present work is derived from the fact that a central
theme in the experimental literature involves distinguishing the
condensate from the normal regions of the trap. Whether there is phase
separation or not \cite{ZSSK06,PLKLH06} the precise nature of the normal
(N) and superfluid (S) phases are all of great interest. In this way one
needs a theory which distinguishes N from S at finite temperatures,
where the excitation gap is no longer a signature of superfluidity.
Previous theoretical approaches, based on the BdG
\cite{Kinnunen,Machida2} and local density approximation (LDA)
\cite{PWY05,SM06,HS06,PS05a} schemes have emphasized $T=0$, albeit
without reaching any clear consensus.  The inclusion of finite $T$ for
the LDA case has been introduced within the same formalism we use here
\cite{YD05}, but without separating the condensed and noncondensed pair
contributions.  In addition the application of BdG to $T \neq 0$ is
viewed as problematic because it does not incorporate noncondensed
pairs \cite{Chen2,Kosztin1,PSP03}.  At the same time, it should be
stressed that this BdG approach \cite{Kinnunen,Machida2} is most likely
the appropriate way to get a full picture of the $T=0$ phase.

The formalism used in this paper was outlined earlier \cite{CCHL06}.
Here we incorporate trap effects by use of the LDA.  We adopt a
one-channel approach since the $^6$Li resonances studied thus far are
broad and consider a Fermi gas of two spin species with kinetic energy
$\epsilon_\mathbf{k} = \hbar^2 k^2/2m$ subject to an attractive contact
potential ($U<0$) between the different spin states.  We define $\delta
n = n_\uparrow -n_\downarrow >0$, where $n= n_\uparrow +n_\downarrow$ is
the total atomic density.  Importantly, we include \cite{JS2}
noncondensed pairs at general $T$.  The $T$-matrix or noncondensed pair
propagator, is $t(Q)= U/[1+U\chi(Q)]$, where $\chi(Q)$ is the pair
susceptibility discussed earlier which depends self consistently on the
fermionic excitation gap $\Delta$.  The presence of pairing correlations
means that $\Delta^2$ contains two additive contributions from the
condensed ($\Delta_{sc}^2$) and noncondensed pairs ($\Delta^2_{pg}$).
In the superfluid phase, we have $1+U\chi(0)=0$, equivalent to
$\mu_{pair}=0$, the BEC condition of the pairs. As a consequence, the
equations become simpler below $T_c$ and we may expand the $T$-matrix to
arrive at a characteristic frequency $\Omegaq$ which characterizes the
dispersion of the noncondensed pairs.

We now summarize the self-consistent equations \cite{JS2,JS5}, in the
presence of a spherical trap, treated at the level of LDA with trap
potential $V(r)=\frac{1}{2}m\omega^2 r^2$.  $T_c$ is defined as the
highest temperature at which the self-consistent equations are satisfied
precisely at the center.  At a temperature $T <T_c$ the superfluid
region extends to a finite radius $R_{sc}$. The particles outside this
radius are in a normal state, with or without a pseudogap.

\begin{figure*}
%\centerline{\includegraphics[height=2.56in,clip]{Ia0_nocap_2.eps}
%~  \includegraphics[height=2.6in,clip]{col_Ia0_2.eps}}
\centerline{\includegraphics[height=2.6in,clip]{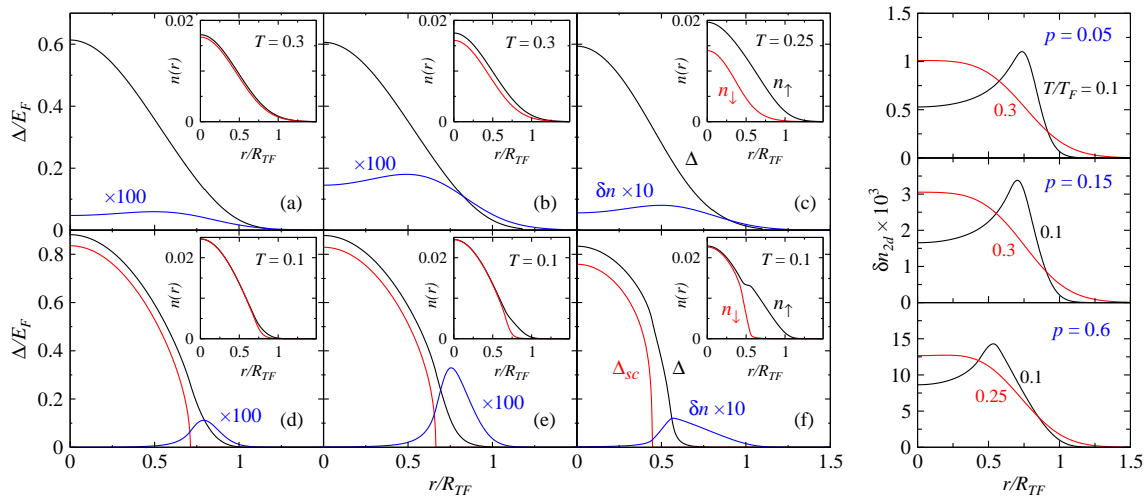}}
\caption{Same as Fig.~\ref{fig:1} but at unitarity. From left to right,
  $p=0.05$, $0.15$, and $0.6$, and $T_c/T_F \approx 0.27$, $0.27$, and
  $0.23$, respectively. For the upper row, $T/T_F = 0.3$, $0.3$, and
  $0.25$, respectively, and for the lower row $T/T_F =0.1$. Shown on the
  far right is $\delta n_{2d} (r)$. }
\label{fig:2}
\end{figure*}

The generalized local gap equation is given by
\begin{equation}
  \frac{m}{4\pi a}= \sum_{\mathbf{k}} \left[\frac{1} {2\ek} - \frac{1-
      2\bar{f}(\Ek)}{\Ek} \right] + Z\mu_{pair},
\label{gap_eq_trap}
\end{equation}
where $\mu_{pair}(r) = 0$ in the superfluid region $r\le R_{sc}$, and
must be solved for self-consistently at larger radii.  The quantity $Z$
is the inverse residue of the $T$-matrix \cite{CCHL06}.  For convenience
we write $ \bar{f}(x) \equiv [f(x+h)+ f(x-h)]/2 $ where $f(x)$ is the
Fermi distribution function.  Here we set $\hbar=1$, and use $ m/4\pi a=
1/U + \sum_{\mathbf{k}}(2\epsilon_{\mathbf{k}})^{-1}$ to regularize the
contact potential, where $a$ is the two-body $s$-wave scattering length.
The dispersion is given by $\Ek = \sqrt{[\ek-\mu(r)]^2 +\Delta^2}$, with
$\mu(r)=(\mu_\uparrow+\mu_\downarrow)/2 - V(r)$. We also define the
r-independent parameter $h = (\mu_\uparrow - \mu_\downarrow)/2$.  Since
$\delta n \ge 0$, we always have $h > 0$.  More generally, $\mu_\sigma$
is the chemical potential for spin $\sigma$ at the trap center.

The local pseudogap contribution (present only at
$ T \neq 0$) to $\Delta^2(T) = \Delta_{sc}^2(T) +
\Delta_{pg}^2(T)$ is given by
\begin{equation}
\Delta_{pg}^2=\frac{1}{Z} \sum_{\bf q}\, b(\Omegaq -\mu_{pair})\,\,.
\label{eq:1}
\end{equation}
where $b(x)$ is the usual Bose distribution function.
The density of particles at radius $r$ can be written as
\begin{equation}
n_\sigma(r) = \sumk [\uk^2f(E_{\mathbf{k}\sigma})
+ \vk^2f(-E_{\mathbf{k}\bar{\sigma}})]\,,
\label{number_eq_trap:above}
\end{equation}
which depends on the coherence factors $u_\mathbf{k}^2, v_\mathbf{k}^2 =
(1\pm \xi_\mathbf{k}/\Ek)/2$ with $\xi_\mathbf{k} = \ek - \mu(r)$, and
$E_{\mathbf{k}\uparrow}=-h+\Ek$ ,and $E_{\mathbf{k}\downarrow}=h+\Ek$.
The total number of particles and the polarization are respectively
given by
%$N_\sigma(r) = \int d^3 r n_\sigma(r)$,
\begin{eqnarray}
N_\sigma(r) &=& \int d^3 r\, n_\sigma(r), \quad
N=N_\uparrow + N_\downarrow, \\
p &=& (N_\uparrow +N_\downarrow)/N.
\end{eqnarray}
Our calculations proceed by numerically solving the self-consistent
equations. Here we use the $N$ and $p$ as input; these are the control
parameters in experiment.

Figure \ref{fig:1} shows the behavior of the various gap parameters and
the majority and minority spin components as a function of radius in the
trap, for the case of a near-BEC system with $1/k_Fa = 1.5$. The upper
panels are for the normal phase and the lower panels are in the
superfluid state.  We present results for three different polarizations
and focus first on the lower panels where there are two distinct
components to the gap $\Delta_{sc}$ and $\Delta_{pg}$. The two gap
functions, $\Delta_{sc}$ and the total gap $\Delta$, are plotted vs $r$
along with the difference density for up and down spins, or
alternatively, the polarization. We overlay these plots into order to
show clearly what are the contributions to the polarization from the
condensate (I), where $\Delta_{sc} \neq 0$, the correlated, but normal
mixed region (II), where $\Delta_{sc} =0$, but $\Delta \neq 0$, and
non-interacting Fermi gas(s) regime (III), where $ \Delta =0$.

It can be seen that there is very little polarization present in the
condensate (I) which appears below $T_c$, as has been inferred experimentally
\cite{ZSSK06,PLKLH06,ZSSK206}.  Rather the bulk of the polarization is
present in the correlated, but normal region (II) in which there is a
finite excitation gap $\Delta$, but vanishing $\Delta_{sc}$.  In region
III at even larger radii, $\Delta$ is essentially zero and region III
is predominantly composed of the majority spin component. In this regime,
one expects the cloud wing-shape to be that of a non-interacting Fermi
gas, and this provides the basis for a reasonable thermometry
\cite{ZSSK206}.  
As $T$ is lowered the noncondensed pairs in Region II will be
converted into the condensate, thereby merging Regions I and II.

Because we have not yet incorporated the $q_0 \neq 0$ correlations of
the FFLO state, in Figs.~\ref{fig:1} and \ref{fig:2} the largest of 3
values of $p$ used is associated with an instability at the very edge of
the minority cloud.  Nevertheless, since $n_\downarrow$ essentially
vanishes there, this is expected to have very little qualitative effect
on our results.

\begin{figure*}
%\centerline{\includegraphics[width=4.3in,clip]{Iap3_nocap_2.eps}
%\centerline{\includegraphics[height=2.6in,clip]{Iap3_nocap_2.eps}
%~  \includegraphics[height=2.6in,clip]{col_Iap3_2.eps}}
\centerline{\includegraphics[height=2.6in,clip]{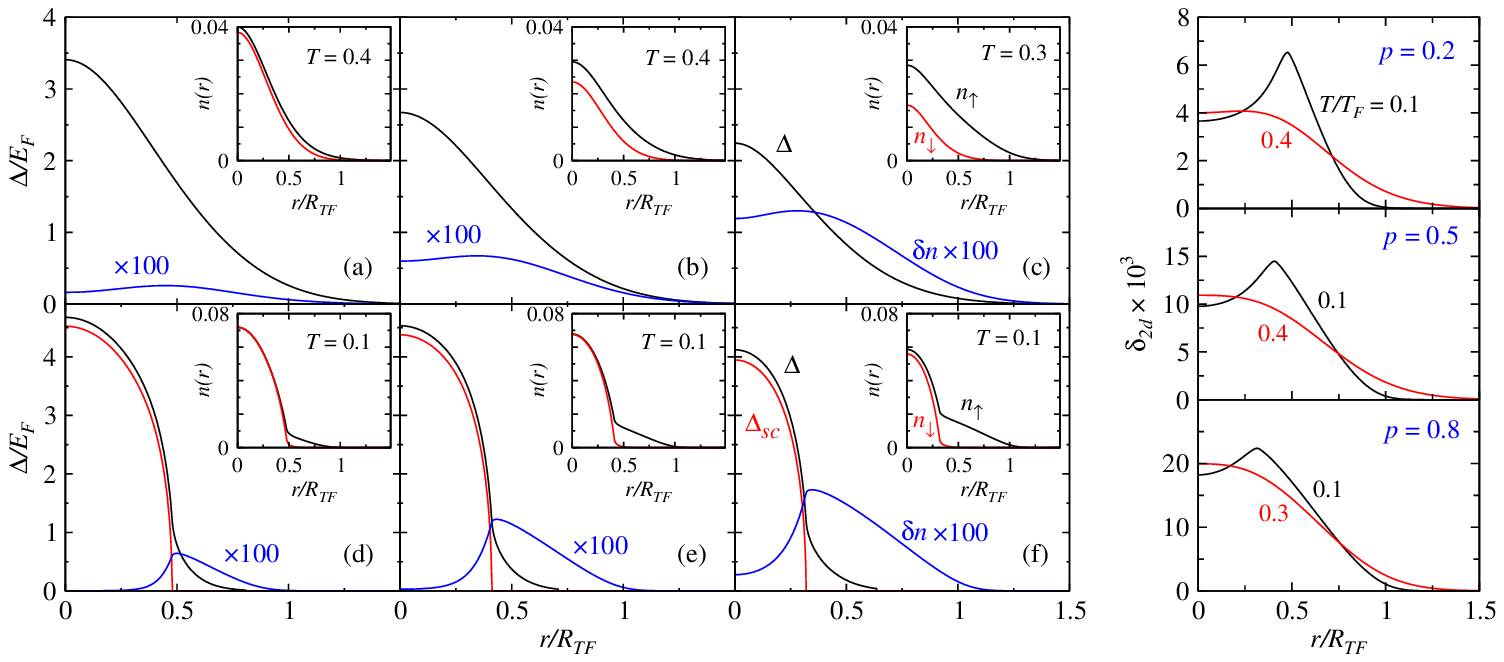}}
\caption{Same as Fig.~\ref{fig:1} but for $1/k_Fa=3$.
  From left to right, $p=0.2$, $0.5$, and $0.8$, and $T_c/T_F\approx
  0.39$, $0.36$, and $0.28$, respectively. For the upper row, $T/T_F =
  0.4$, $0.4$, and $0.3$, respectively, and for the lower row $T/T_F
  =0.1$. Shown on the far right is $\delta n_{2d} (r)$.  }
\label{fig:3}
\end{figure*}

The insets in the lower panel show the density profiles for the majority
(in black) and minority (in red) component.  Qualitatively similar to
what has been observed experimentally \cite{ZSSK206}, a small ``kink''
in the majority is present at the radius at which the condensate ends.
The minority component contains essentially only a condensate central
peak with a very weak bi-modal structure.
The upper panel shows the behavior in the normal state, where the
condensate $\Delta_{sc} =0$.  Nevertheless, it can be seen that an
excitation gap $\Delta$ is present throughout the cloud. In this way the
particle profiles do not correspond to those of a non-interacting gas,
and the polarization is rather evenly distributed at all radii in the
cloud.  It may also be seen from these insets that as the system varies
from above to below $T_c$, the profile of the minority component
contracts into the center of the trap, as observed \cite{ZSSK206}.

We present comparable figures for the unitary case in Fig.~\ref{fig:2}.
Most of the observations made above for the near-BEC case obtain at
unitarity as well. Here, one can see from the insets, however, that the
kink in the majority profile is less apparent.  Finally, we turn to
Fig.~\ref{fig:3} where the counterpart plots are presented for the BEC
regime with $1/k_F a = 3.0$.  From the right column, it may be seen that
polarization penetrates down to the trap center, when the overall
polarization $p$ is high. This is in agreement with the expectations
from the homogeneous case \cite{CCHL06}.

It should be stressed that our calculations indicate that dramatic
changes in the shape of the density profiles do not occur until $T$ is
substantially lower than $T_c$. This may explain why the experimentally
observed $T_c$ values are less \cite{ZSSK206} than those we compute.
Also important is the fact that the mixed normal phase we find here
(Region II) is not related to that introduced at $T=0$ in other
theoretical work \cite{SM06,HS06}.  The noncondensed pair contribution
we consider has no counterpart at $T=0$.  Moreover, there is no evidence
from $ T=0$ BdG investigations \cite{Kinnunen,Machida2} for such a mixed
normal phase, although whether it is consistent with the FFLO state in a
trap bears further investigation.

Essentially all the qualitative observations reported in this paper
correspond to their counterparts in Ref.~\cite{ZSSK206} with one
exception.  In Ref.~\cite{ZSSK206}, it is claimed that only regions I
and III are present in the near-BEC regime, whereas, we find all three
regions appear, just as in the unitary case.  Region II corresponds
quite naturally to the presence of noncondensed pairs, expected at
finite $T$ in the near-BEC regime.  At a more quantitative level, our
results at unitarity may change somewhat when we include FFLO condensate
contributions, and associated higher values of $p$.  While we do not
find evidence for sharp phase separation as reported in
Ref.~\cite{PLKLH06}, after column integration of Figs.
\ref{fig:1}-\ref{fig:3}, a double peaked structure emerges for the
difference profile $\delta n_{2d}$ at low $T$, as has been claimed
experimentally \cite{PLKLH06,ZSSK06}.  Importantly, as stressed in
Ref.~\cite{ZSSK206}, and, as is consistent with our longstanding
viewpoint \cite{ourreview}, these studies show that a significant
fraction of the normal region in the trap contains strong interactions
between the two spin states which we associate with the presence of
noncondensed pairs and related fermionic excitation (pseudo)gap.

This work was supported by NSF-MRSEC Grant No.~DMR-0213745. We thank
M.W. Zwierlein for useful communications.

%\vskip -2ex

\bibliographystyle{apsrev}

%\bibliography{Review2}

\end{document}